\documentstyle[12pt,aaspp4]{article}
\lefthead{Hashimoto et al.}
\righthead{Influence of Environments on Star Formation}

\begin{document}

\title{THE INFLUENCE OF ENVIRONMENT ON THE STAR FORMATION RATES OF
GALAXIES} 

\author{Yasuhiro Hashimoto and Augustus Oemler, Jr.}
\affil{Carnegie Observatories, 813 Santa Barbara St., Pasadena, CA
91101 USA; and Department of Astronomy, Yale University, P.O. Box
208101, New Haven, CT 06520-8101 USA; hashimot@astro.yale.edu,
oemler@ociw.edu}

\author{Huan Lin} 
\affil{Department of Astronomy, University of Toronto,
    Toronto, Ontario, M5S 3H8, Canada; lin@astro.utoronto.ca}
\and
\author{Douglas L. Tucker}
\affil{Fermilab, MS 127, PO Box 500,
    Batavia, IL 60510 USA; dtucker@fnal.gov}

\begin{abstract}

We have used a sample of 15749 galaxies taken from the Las Campanas
Redshift Survey (Shectman et al. 1996) to investigate the effects of
environment on the rate of star formation in galaxies. 
For each galaxy we derive 
a measure of star formation
rate (SFR) based on the strength of the [OII] emission line, 
and a measure of galactic structure based on the central
concentration of the galaxy light, which is used to decouple the
effect of ``morphology-environment'' relation from the SFR.
Galactic
environment is characterized {\em both} by the three-space local density of 
galaxies and
by membership in groups and clusters.

The size and homogeneity of this data set allows us to sample, for the
first time, the entire range of galactic environment, from the lowest
density voids to the richest clusters, in a uniform
manner. Thus, we could expand our research from
the conventional cluster vs. field comparison to
a new ``general''
environmental investigation by decoupling the local galaxy density
from the membership in associations.
This decoupling is very crucial  for constraining the physical
processes responsible for the observed environmental dependencies of
star formation.
On the other hand, the use of an automatic measure of galactic structure
(concentration index), rather than Hubble type which is subjective and
star formation-contaminated estimate of galactic morphologies, allows us to 
cleanly
separate the morphological component from the SFR vs. environment
relationship.

We find that, when cluster/field comparison is made, cluster galaxies exhibit
{\it reduced} star formation {\em for the same concentration index}.
This result supports several previous Hubble type-based studies
reporting similar suppressions of star formation among
cluster galaxies for the same Hubble type.
We did not find any qualitatively different responses to environments
between early and late type spirals, which were also previously reported.
On the other hand, a further division of clusters by ``richness''  reveals
a new possible excitation of starbursts in groups and poor clusters.

Meanwhile, a more general environmental investigation
shows that
the star formation rate of galaxies of a given
concentration index is sensitive to local galaxy density and shows
a continuous correlation with the local density,
in such a way that galaxies show higher levels of star formation
in low density than in high density environments.
Interestingly, this trend is also observed both inside and outside of clusters,
implying that physical processes responsible for this
correlation might not operate intrinsically in the cluster environment.  
Furthermore, a more complex facet of the dependence of SFR on local density is
also revealed;
galaxies with differing levels of star formation  appear to
respond differently to the local density.
Low levels of star formation, corresponding to
those expected in normal members of the Hubble sequence, are more
sensitive to environment inside than outside of clusters. In contrast,
high levels of star formation, identified as ``starbursts'', are at
least as sensitive to local density in the field as in clusters.

We conclude that at least two separate processes are responsible for
the environmental sensitivity of the SFR, and tentatively identify
gas removal processes as responsible for the variation with density of
the SFR of normal galaxies, and galaxy-galaxy interactions as
responsible for the prevalence of starbursts in intermediate density
environments.

\end{abstract}


\section{INTRODUCTION}

Many lines of evidence, accumulated over the past several decades,
have made it abundantly clear that there has been substantial
evolution of the properties and populations of galaxies during recent
epochs ($z \leq 1.0$). Such evolution is seen both in clusters
(Butcher \& Oemler 1984; Dressler et al. 1994),
in groups (Allington-Smith et al. 1993) and in the general ``field''
population (Lilly et al. 1996, Glazebrook et al. 1995). Although it
is possible that some portion of the evolutionary changes observed are
due to causes internal to the individual galaxies, there are reasons
for suspecting that much of the evolution is driven by external forces
in the galaxies' environment.
 
Firstly, there is some evidence which suggests that the rate at which
galaxies of a given type have evolved varies substantially with
environment (Allington-Smith et al. 1993). Secondly, the profound
changes in the global properties of galaxies which have occurred
during recent epochs are difficult to understand using only the
processes occurring within an undisturbed, isolated galaxy. Finally,
the large systematic variations in galaxy populations with environment
{\em require} that environment affected the
properties of galaxies no earlier than the epoch of their formation.

A number of modes of interaction of galaxies and their surroundings
are known which can be expected to cause significant changes in the
properties of galaxies over time, and at least some examples of such
processes in action have been discovered in low and intermediate
redshift galaxy populations. Among these processes are galaxy-galaxy
mergers (e.g. Barnes \& Hernquist 1991), 
tidal interactions between galaxies and
surrounding masses (e.g. Byrd \& Valtonen 1990), 
and gas-phase interactions between the
intragalactic and intergalactic media 
(e.g. Gunn \& Gott 1972; van den Bergh 1976; Dressler \& Gunn 1983).
Unfortunately, however easily one may enumerate processes to
drive galaxy evolution, and however plausible such mechanisms may be,
there exists little evidence demonstrating that any one of these
processes is, in fact, responsible for driving galaxy evolution. Most
of these processes act over an extended period of time, while
observations of any population of galaxies give one only a snapshot of
one instant in its history. Also, observations at intermediate and
high redshifts, which have provided most of the evidence for galaxy
evolution, cannot easily provide the detailed information which is
needed to elucidate subtle and complicated processes.

However, any means by which galaxies interact with their surroundings
should operate today as well as at earlier times, and, therefore,
observations of nearby populations should be an effective way of
understanding them. In addition to direct evidence for the occurrence
of interactions capable of driving galaxy evolution, the local
universe holds clues about the nature of those phenomena in the form
of variations of galaxy properties with environment, which must be the
result of past interactions. The most well-established of such
variation is the morphology-density relation (Dressler 1980), which
appears to hold not only in clusters, but also in the field 
(Bhavsar 1981; de Souza et al. 1982; Postman
\& Geller 1984; Giovanelli \& Haynes 1986; Tully 1988). 
However, another, perhaps
equally important clue for the understanding of the origin and the
evolution of galaxies can be obtained by the study of the influence of
environment on star formation of galaxies, since star formation is
both a fundamental galactic parameter and a driver of galaxy
evolution.

There have been many studies investigating environmental influences on
star formation. However, these previous studies have been limited to
cluster vs. field comparisons (or similar membership comparisons)
which compare star formation in cluster
galaxies to that in a field control sample. Unfortunately, these
studies have produced conflicting results. Some have suggested a {\em
reduced} star formation rate (SFR) in cluster galaxies with respect to
field galaxies of the same morphological type (e.g., Gisler 1978;
Kennicutt 1983; Dressler, Thompson \& Shectman 1985). However, other
studies suggest a similar or {\em higher} SFR in cluster spirals with
respect to the field sample (e.g., Kennicutt et al.
1984; Gavazzi \& Jaffe 1985). Others report qualitatively
differential responses to environment between early and late type
galaxies (e.g. Moss \& Whittle 1993).

Some of the inconsistency between previous cluster/field studies can be
traced to two facts. Firstly, these studies use small samples for
both cluster and field subsets (at most, on the order of 10$^{2}$
galaxies in each subset). With samples of this size it is difficult to
make a statistically sound comparison, particularly after binning
galaxies by Hubble type. Also, the field samples are usually selected
from existing bright galaxy catalogs, or from ``pencil beam'' studies,
which, despite a careful effort to select a fairly normal cross
section of galaxies, may contain galaxies from a wide variety of
environments. Thus ``field'' samples may contain galaxies within
loose groups or the periphery of clusters which dilute the contrast
with the cluster samples. Secondly, most of the previous cluster/field
studies were forced to combine multiple data sets with heterogeneous
characteristics, such as different Hubble classifications by
different observers, varying image quality, different star formation
measures with different sensitivity, and even different selection
criteria of sample objects, all of which can cause spurious results.

It is clear that a new cluster/field comparison using
a large number of galaxies studied in a consistent manner is very much needed.
However, even a cluster vs field comparison free of these problems is
not sufficient. The cluster/field studies (or any studies comparing
subsets which are selected on the basis of membership in galaxy
associations) have a fundamental limitation for understanding the
mechanisms responsible for the environmental influences on star
formation. In such studies, the inability to decouple very local
environment, as characterized, for example, by local galaxy density,
from more global environments, such as membership in a cluster,
prevents us from differentiating between mechanisms specific to each
of these classes of environment. The significance of the distinction
between local and broader environments has been a matter of much
contention in cluster studies (see, for example, Dressler 1980 vs.
Whitmore \& Gilmore 1991)

A further, equally serious, problem with previous studies comes from
the nature of the Hubble type itself, which has been used for the
normalization of star formation rates over the broad range of galaxy
types. The Hubble type is determined by multiple characteristics of a
galaxy, one being the resolution of spiral arms. However, the
resolution of spiral arms is, in practice, determined largely by the
star formation activity in the arms. Thus, systematic variations in
the star formation may cause a systematic shift in the Hubble
type. When Hubble type is used to normalize star formation rates, this
shift results in a serious reduction in the sensitivity of the
measurement of varying star formation rates. To avoid this, we need
to characterize the galaxies not by Hubble type, but rather by a
physical parameter which is more independent of star formation. (The
problem of the Hubble system is further discussed in \S5.)

In this paper, we present our first attempt to answer questions about
general environmental effects on the star-forming properties of
galaxies in the local universe, taking advantage of the very large and
homogeneous data set available from the Las Campanas Redshift Survey
(LCRS; \cite{sh96}). This data set consists of a large number 
of galaxies inhabiting the entire range of galactic environments, from
the sparsest field to the densest clusters, thus allowing us to study
environmental variations without combing multiple data sets with
inhomogeneous characteristics. Furthermore, we can also extend our
research from the traditional cluster/field comparison to more
``general'' environmental study by, for the first time in
investigations of star formation properties, decoupling the local galaxy
density from the membership in associations. Finally, to minimize the
problems with the use of Hubble types mentioned above, we have used an
automatically measured concentration index as a star formation
baseline.

The outline of this paper is as follows. \S 2 briefly describes the
dataset used. In \S 3 we describe the spectroscopic measures of star
formation, and in \S 4 we discuss the method of analysis. Results are
in \S 5.

\section{DATA}

Here we briefly describe our survey parameters; the reader is referred
to Shectman et al.\ (1996) for further details. The LCRS consists of
26418 galaxies, with a mean redshift $z = 0.1$, and a depth of about $
z = 0.2$. The survey galaxies were selected, using isophotal and
central magnitude criteria, from CCD-based photometry in a ``hybrid''
Kron-Cousin $R$ band. This photometry was obtained from driftscans on
the Las Campanas 1m Swope Telescope. The survey covers over 700 square
degrees in six $1.5^{\circ} \times 80^{\circ}$ ``slices'' in the North
and South galactic caps. Every slice consists of over 50 $1.5^{\circ}
\times 1.5^{\circ}$ spectroscopic fields, each containing a maximum of
50 or 112 galaxies.  The first 20\% of the redshifts were obtained
using a 50-object fiber-optic spectrograph. The nominal isophotal
magnitude limits of the spectroscopic sample were $16.0 \leq R \leq
17.3$, and an additional central magnitude limit excluded the 20\% of
galaxies of lowest central surface brightness. The rest of the
redshifts were obtained with a 112-object fiber system, with isophotal
limits of $15.0 \leq R \leq $17.7, and exclusion of the 5-10\% of
galaxies with lowest central surface brightness.
The shape of the luminosity function of the LCRS is consistent with that
of other redshift surveys (Lin et al. 1996). 

The spectra were obtained with the multi-fiber spectrograph and
Reticon detector mounted on the du Pont 2.5m telescope at Las Campanas
Observatory. Each spectrum was flat fielded, wavelength calibrated,
and sky subtracted. The spectra have a wavelength range of 3350-6750
\AA, with a resolution of $\sim$ 5 \AA \ and a pixel scale of $\sim$ 3
\AA. The average signal-to-noise in the continuum around the Balmer
absorption lines is 8 to 9.

\section{SPECTROSCOPIC MEASURES}

In order to quantify the star formation properties of LCRS galaxies,
we have measured the equivalent width (EW) of [OII]$ \lambda$3727,
[OIII]$ \lambda$5007, and H$\beta$ in LCRS spectra. Conventionally,
H$\alpha$ has proved to be the best optical indicator of the massive
star formation rate (e.g. Kennicutt 1983). However, H$\alpha$ is
inaccessible in many of our spectra, due to the redshift range and
spectral coverage of the survey. Several workers have used the EW of
the [OII] $\lambda$3727 doublet or of H$\beta$ as a star formation
index for distant galaxies (e.g. Dressler \& Gunn 1982; 
Dressler et al. 1985; Peterson et al. 1986; Broadhurst et al. 1988; 
Lavery \&
Henry 1988; Colless et al. 1990). Gallagher et al. (1989) have
derived an approximate [OII] versus star formation rate (SFR)
calibration from observations of [OII] and H$\beta$ in nearby blue
galaxies. A direct comparison to EW(H$\alpha +$ NII) (Kennicutt 1992)
showed that EW(H$\beta$) and EW(OIII $\lambda$5007) can serve as good
substitutes for star forming indicators in strong emission galaxies
(those with EW(H$\alpha +$ NII) $\geq 60 $\AA, and EW(H$\beta$) $\geq
5 $ \AA), while EW(OII) is a good indicator of star formation for all
emission strengths.

The equivalent widths are measured automatically by integrating the
signal above or below the local continuum outward from the center of
the line until reaching the continuum level. The local continuum is
determined by fitting a third order polynomial over the 350 \AA\ on
either side of the line, excluding the line itself and nearby sky
lines. The algorithm iterates the fitting 3 times while excluding the
points outside 2 sigma of the continuum. The equivalent width
uncertainties are calculated using Poisson statistics, the local noise
in the continuum, and standard propagation of errors. The mean errors
in the measurements of EW(H$\beta$), EW(OII), and EW(OIII) are 1.8
\AA, 2.2 \AA, and 2.1 \AA\  respectively. Those galaxies with a
continuum signal-to-noise ratio S/N $ <$ 6 within the 25 \AA\  window
centered on each line are excluded from the analysis. The number of
galaxies remaining for EW measurements after this S/N cut depends on
the line measured; it is 18875 for the H$\beta$ line, 16377 for the
[OII] line and 17351 for the [OIII] line.

The fibers in the du Pont multi-object spectrograph subtend a circle
3\farcs5 in diameter. For galaxies with recessional velocities
between 15,000 and 40,000 km s$^{-1}$ (a velocity cut which was used
for the analysis, as discussed below), this diameter corresponds to a
projected circle of 5 $\sim$ 10 kpc (H$_{0}$ = 50 km s$^{-1}$
Mpc$^{-1}$, q$_{0}$ = 0.5). This is smaller than the total size of
the typical galaxy, but {\em much} larger than the nuclear regions, and
considerably larger than the bulges of all but the most
bulge-dominated galaxy. The result of this undersampling of the disks
will be a small systematic underestimate of star formation rates in
the earliest spirals. However, such a systematic shift will have no affect on
any of the analysis presented later in this paper.

\section{ENVIRONMENTAL PARAMETERS}

\subsection{Local Galaxy Density}

To characterize the environment of LCRS galaxies, we calculate the
local galaxy density, $\rho$, around each of the 26418 galaxies using
a nearest neighbor technique. For each galaxy, we take the local
galaxy density to be
\begin{equation}
\rho=\frac{3}{\frac{4}{3}\pi D^{3}} \ ,
\end{equation}
where $D$ is the three-dimensional redshift-space distance from the
galaxy to its third nearest neighbor. Note that this measure of
galaxy density uses a three-space distance to nearest neighbors.
Since the radial component of this distance is derived form the
galaxy's redshift, the effect of the peculiar velocities will be
spread out the neighboring galaxies along the line of sight and thus
to cause a systematic underestimate of the density in the densest
regions. However, in even the densest regions of the LCRS sample, we
calculate the underestimate of $\rho$ caused by the peculiar
velocities to be typically less than $\sim$ 20 \%, which is smaller
than the width of the bins of $\rho$ that we used for the analysis.
Thus, the effect is negligible for the purpose of this study.
 
The effect of the variation of the survey selection function at
different redshifts is removed by introducing a weight
\begin{equation}
     w(z_{i})= \frac{1}{S(z_{i})}
\end{equation}
for each galaxy $i$, where
\begin{equation}
 S(z_{i}) =
 \int^{min[M_{max}(z_{i}),M_{2}]}_{max[M_{min}(z_{i}),M_{1}]}\phi(M)dM
 \left/
   \int^{M_{2}}_{M_{1}} \phi(M)dM
 \right. \ ,
\end{equation}
$M_{1}, M_{2}$ are the absolute magnitude limits in which we are
interested, and $ M_{max}(z_{i})$ and $ M_{min}(z_{i})$ are the
absolute magnitude limits, at the redshift of galaxy $i$,
corresponding to the apparent magnitude limits for the field
containing galaxy $i$. We describe the differential luminosity
function $\phi$ by a Schechter function with parameters $\phi^{*} =
0.019 \ h^3$~Mpc$^{-3}$, $M_{R}^{*} = -20.29 + 5 \log h,$ and $ \alpha
= -0.70 $ (Lin et al. 1996), which we assume to be invariant with
redshift.
 
In addition, another weight $W_i$ is calculated for each galaxy $i$ to
take account of the field-to-field spectroscopic sampling variations.
The spectroscopic completeness of a field decreases as the projected
density of galaxies in the field increases, since each spectroscopic
field is observed only once, using a maximum of 50 or 112
fibers. Since galaxies in denser regions were selected randomly for
spectroscopy from among all galaxies meeting the photometric criteria,
this effect is corrected by setting $ W_{i}$ to be the inverse of the
fraction of spectroscopically observed galaxies in the field
containing galaxy $i$. (Additionally, small effects from magnitude
errors, apparent magnitude and surface brightness incompletenesses,
and central surface brightness selection are also included in the
calculated $W_{i}$.  Further detailed discussions of these weights and
corrections are given in Lin et al. 1996.)
 
Now, the corrected local galaxy density $\rho$ around a galaxy $i$
becomes
\begin{equation}
\rho_{i}=\frac{\Sigma ^{3}_{j=1}w(z_{j})W_{j}}{\frac{4}{3}\pi D^{3}} \ ,
\end{equation}
where $j$ represents the rank of the nearest neighbors from galaxy $i$.

After removing objects too close to the LCRS survey spatial boundary,
an additional conservative velocity boundary (from 15000 to 40000 km
s$^{-1}$) was set in order to further minimize the uncertainties in
the density estimate, by allowing the use of only a relatively
constant selection function. The number of galaxies remaining after
the velocity and spatial boundary cuts is 10536.

\subsection{Membership}

As a second environmental parameter, cluster or rich-group membership
was determined for each galaxy in the LCRS. Cluster and rich group
galaxies are defined by the three-dimensional ``friends-of-friends''
group identification algorithm (Huchra \& Geller 1982). The algorithm
finds all pairs within a projected separation $D_{L}$, and within a
line of sight velocity difference $V_{L}$. Pairs with a member in
common are linked into a single group. This linking makes the
membership more sensitive to the environment of larger scale than the
local density parameter defined in \S4.1. The selection parameters
$D_{L}$ and $V_{L}$ are scaled to account for the magnitude limit of
the LCRS survey, and defined as $D_{L}=S_{L}D_{0}$\ and $
V_{L}=S_{L}V_{0}$. Here the linking scale S$_{L}$ is calculated by
\begin{equation}
S_{L} = \Biggl{[}\frac{\rho^{'}(d_{f})}{\rho^{'}(d)}\Biggr{]}^{1/3} \ ,
\end{equation}
where $\rho^{'}(d)$ is the galaxy number density, at the mean comoving
distance $d$ of the galaxy pair in question, for a homogeneous sample
that has the same selection function as the LCRS. In other words,
$\rho^{'}(d)$ is equivalent to the unnormalized galaxy selection
function. 

The distance $d_{f}$ is the fiducial comoving distance at redshift
z$_{f}$ (we chose cz$_{f}$=30000 km s$^{-1}$) at which we define
$D_{0}$ and $V_{0}$. The density enhancement contour surrounding each
group is related to $D_{0}$ by
\begin{equation}
\frac{\Delta\rho}{\rho}=\frac{3}{4{\pi}D_{0}^{3}\rho^{'}(d_{f})}-1
\end{equation}
The values of $D_{0}$ (or $\Delta\rho/\rho$) and $V_{0}$ used are
taken from the LCRS group catalog (Tucker 1994; Tucker et al. 1998),
and are $D_{0}$= 0.72 $h^{-1}$ Mpc (or $\Delta\rho/\rho$ =80) and
$V_{0}$ = 500 km s$^{-1}$, which are determined by several
semi-quantitative constraints similar to those used in Huchra \&
Geller (1982), to avoid biasing the velocity dispersions of groups,
and to optimize the number of interlopers.

\section{RESULTS}

\subsection{Emission Properties of the Sample}

Figure 1 presents the equivalent widths of [OII]$ \lambda$3727 and
H$\beta$ for 15749 LCRS galaxies; emission is represented by positive
values. The majority (13951) of galaxies show negligible or weak
emission, which we define to mean that EW(OII) $<$ 20 \AA\ and
EW(H$\beta$) $<$ 5 \AA. This is the range expected of normal galaxies
of types E to Sc; Kennicutt \& Kent (1983) and Romanishin (1990) have
shown that such galaxies have EW(H$\alpha$ + NII) $\leq$ 50 \AA,
equivalent to EW(OII) $\leq$ 20 \AA, assuming EW(OII) = 0.4
EW(H$\alpha$) (Kennicutt 1992). However, a rather significant
fraction (1798 or $\sim$10\%) of galaxies shows strong [OII] or
H$\beta$ emission (EW(OII) $\geq$ 20 \AA\ or EW(H$\beta$) $\geq$ 5
\AA), with a pattern of line strengths consistent with strong star
formation activity. A typical spectrum of a strong star forming
galaxy is shown in Fig. 2(a). It shows strong [OII], [OIII], and
Balmer emission lines, but weak or undetectable [NeV] lines.

There is a small, fairly distinctive population ($\sim$30) of galaxies
with [OII] emission that is weak compared to the strength of H$\beta$;
typically these have EW(H$\beta$) $\sim$ 20 \AA\ and EW(OII) $<$ 10
\AA\ . An example of such a spectrum is presented in Figure 2(b).
These galaxies show strong broad Balmer emission lines, strong [OIII]
emission, but weak emission in [OII]; most are Seyfert 1 galaxies
(Kennicutt 1992).

Another group of $\sim$40 galaxies shows relatively strong [OII]
emission, independent of Balmer line strength (most of these occupy
the upper left corner of Fig. 1). A typical spectrum (Fig. 2(c))
shows that these galaxies have strong [OIII] emission with respect to
Balmer emission, in addition to strong [OII] emission. These spectra
are also often accompanied by [NeV]$\lambda$3425 emissions and are
suspected to be mostly Seyfert 2's or LINER's (Kennicutt 1992).

\subsubsection{Emission Classes}

For the analysis which follows, it is convenient to divide the
galaxies into classes of emission line strength. We use the [OII]
equivalent width, and define three classes: {\em no emission} or NEM,
for which EW(OII) $< $ 5 \AA, {\em weak emission} or WEM, for which
5 \AA\ $ \leq $ EW(OII) $< $ 20 \AA, and {\em strong emission}, or
SEM, for which EW(OII) $\geq $ 20 \AA.

Assuming EW(OII) = 0.4 EW(H$\alpha$) (Kennicutt 1992), the EW(OII) =
20 \AA \ upper boundary of the WEM class corresponds to EW(H$\alpha$)
= 50 \AA. This boundary is where Kennicutt \& Kent (1983) found the
upper limit of EW(H$\alpha$) for the normal spirals. Thus, it
is plausible that the WEM galaxies are predominantly ``normal''
galaxies, in which star formation is governed by {\em internal}
factors such as gas content and disk kinematics. 

In contrast to these, Kennicutt et. al (1987) found that galaxies with
EW(H$\alpha \ge 50 \AA$) are usually members of close pairs and
suggested that their star formation rates are only weakly correlated
with their internal properties, and much more correlated with external
influence. In fact, many studies suggest that a large fraction of
``starburst'' galaxies are members of interacting systems (e.g.,
Heidmann and Kalloghlian 1973; Wasilewski 1983),
and therefore that interactions of galaxies are one of the important
triggering mechanisms for starbursts. If this is correct, and if
starbursts predominate in the SEM class, the variation with
environment of the fraction of SEM galaxies may reflect environmental
variations in galaxy interraction rates.

\subsubsection{AGNs}

Using EW(OII) as an indicator of the SFR fails for galaxies with
luminous active nuclei (AGNs) (Kennicutt 1992). Thus, excluding AGNs
is desirable, even if the total number of AGNs is minimal,
particularly if AGN activity has a different dependence on environment
than does star formation activity. AGN galaxies fall into two
classes, those with abnormally strong [OII] emission, independent of
Balmer line strength, and those with strong Balmer, and [OIII]
emission but weak emission in [OII]. The former most often are Seyfert
2 or LINER, and the latter tend to be Seyfert 1, though there are
exceptions to this rule (Kennicutt 1992).

Seyfert 1 galaxies are relatively easy to identify. With their weaker
[OII] emission with respect to the H$\beta$ emission, and their broad
Balmer lines, Seyfert 1 galaxies are a distinctive population among
emission galaxies in the EW(OII) vs. EW(H$\beta$) plane. We exclude
33 galaxies from the sample with EW(H$\beta$) $\geq$ 7 \AA\ and
EW(OII) $\leq$\ 15 \AA, all of whom show broad Balmer emission lines.
Identifying Seyfert2's or LINER's is somewhat less reliable using the
EW(OII) vs. EW(H$\beta$) plane alone, without access to the H$\alpha$
line. We did our best by additionally using EW(OIII) and
EW(NeV($\lambda$3425)) to exclude a larger subset of 45 galaxies which
{\it includes} Seyfert2/LINER galaxies, but also probably some
non-AGNs, using the criteria EW(OIII)/EW(H$\beta$) $\geq$ 2 or
EW(NeV($\lambda$3425)) $\geq$ 7 \AA.

\subsection{Morphology}

The goal of this paper is to investigate how star formation rates
within galaxies vary with environment. However, the star formation
rate in galaxies, in the weak star formation regime in particular, is
generally correlated with their morphological type (Kennicutt \& Kent
1983). Moreover, the distribution of morphological type itself is a
function of environment (Dressler 1980). Thus, comparison of star
formation rates among environmental subsets needs to account for any
differences in the morphological distribution among the galaxy
subsets.

Our task is, however, complicated by the fact that the star formation
rate is one of the parameters which {\em define} the morphological
type. In the Hubble System (Sandage 1961), morphological type is
based on three characteristics: bulge-to-disk ratio (B/D), tightness
of spiral arms, and the degree of resolution of spiral arms. It is
widely believed that the tightness of spiral arms is related to the
mass distribution within a galaxy, and therefore to B/D. The degree
of resolution of spiral arms is, on the other hand, strongly affected,
even defined, by the star formation activity in the arms.
Thus, two (rather than one) physical parameters, star formation rate
and mass distribution, map into three parameters characterizing the
Hubble System, and this star formation dependency of the Hubble System
complicates our analysis.

The reason is straightforward to understand. If a particular
environment reduces the SFR in a galaxy, this decrease in SFR will
shift its Hubble type towards an earlier type. Because the galaxy
appears with an earlier Hubble type, we will {\em expect} a lower SFR,
and therefore underestimate the amount by which star formation has
been diminshed. The result is a lowered sensitivity to environmental
changes in SFR. 
Van den Bergh (1976), who introduced the term
`anemic' to refer to galaxies with weak star formation,
in a way recognized this dangerous star formation dependence of the Hubble type.
However, the better cure is using a measure of galaxy morphology 
which is
{\em independent} of SFR. From the earlier discussion of the Hubble
sequence, it is clear that the natural candidate is mass distribution,
but that can only be determined from rotation curves, and is,
therefore, an impractical measure for any large sample of galaxies. A
more practical measure is the light distribution.

We quantify the light distribution of the LCRS galaxies using the
automatically-determined concentration index $C$ (Okamura, Kodaira, \&
Watanabe 1983; Doi, Fukugita, \& Okamura 1993; Abraham et al. 1994),
which measures the intensity-weighted second moment of the galaxies
and compares the flux between the inner ($r$ $<$\ 0.3) and outer ($r$
$<$\ 1) isophote to indicate the degree of light concentration in the
galaxy images. 
Here $r$ is a normalized radius which is constant on an elliptical
isophote and is normalized in such a way that $r$ is unity when the
area within the ellipse is equal to the detection area of a galaxy.
(For further details of the definition, please see Abraham et
al. 1994.)

The concentration index C has been developed as a substitute for
Hubble type, however, we stress that C actually has a significant
advantage over Hubble type for the purpose of investigations of star
formation. In other words, it is as a purer measure of one of the two
physical parameters determining Hubble type that we make use of
it. Concentration not only suits our needs better than Hubble type, it
is also more robust against image degradation, and also easier to be
measured automatically. It is, thus, ideal for a large galaxy survey,
such as the LCRS, where the sample size is $\sim$ 10$^4$ and most of
the galaxies consist of on the order of 10$^2$ resolution elements.
(Image parameters of LCRS galaxies are further discussed in Hashimoto
et al. 1998).

In Figure 3 we present the relation between C and mean [OII]
equivalent width of the galaxies in our sample. This figure shows a
smooth increase in mean EW with decreasing C which parallels the
relation between the Hubble type and EW(H$\alpha$) (e.g. Kennicutt \&
Kent 1983), or EW(OII)(Kennicutt 1992). This is the relationship
which we shall use as a baseline for the comparison of the star
formation rates. Figure 4 shows the distribution of C in each of the
three emission classes. The distribution for the WEM class is more
skewed toward late/irregular type galaxies (smaller C) than is that of
the NEM class, as one would expect. The SEM class, on the other hand,
shows a C distribution roughly identical to that of the WEM class,
suggesting that the influence of mass distribution, or ``galactic
structure'', is minimal here. Thus, differences between the SEM and
WEM classes must be due entirely to factors other than galactic
structure.

\subsection{Correlation with Local Density}
\subsubsection{Density effects in the field}

Figure 5 shows the SEM/WEM and WEM/NEM population ratios as a function
of the local space density for the LCRS sample. Bars are root N error.
{\em The small difference in the C distribution between SEM and WEM shown
in Fig. 4 has been removed in order to ensure that the population
difference with respect to the density does not come from an indirect
result of a correlation of galactic structure with density}. The
correction is made by assigning a weight to each WEM galaxy in a given
C bin so that the sum of the weight of WEM galaxies in that C bin will
be equal to the total {\it number} of SEM galaxies in the same C bin.
All WEM galaxies in one C bin carry exactly the same weight and this
weight, instead of the count, is used for WEM galaxies throughout the
population analyses. A similar correction is made between WEM and NEM
by assigning weights to NEM galaxies to match the C distribution of
the WEM class.

Emission line strengths are sometimes found to be correlated with
galaxy luminosity, in the sense that galaxies of lower luminosity
exhibit stronger emission (e.g. Kennicutt \& Kent 1983; Kennicutt
et. al 1984). This trend, however, is primarily due to the
correlation of absolute magnitude with morphological type: late-type
galaxies tend to be both less luminous and exhibit stronger star
formation. Since we applied the correction using C, this luminosity
bias is mostly removed. As an extra precaution, however, to ensure
that we have no EW biases with respect to redshift, we further remove
any difference in the distributions of absolute R magnitudes between
SEM \& WEM, or WEM \& NEM, by assigning additional weights to WEM or
NEM galaxies, until the shapes of their absolute magnitude
distributions match that of SEM or WEM galaxies, respectively.
(Hereafter, whenever the correction using C is applied, it is always
accompanied by an additional absolute magnitude correction.)

Figure 5 shows the correlation between the local density and the
population ratios of different emission types. Fig. 5a shows the
population ratio of SEM to the WEM emission class versus the local
galaxy space density ($\rho$), while Fig. 5b shows the corresponding
ratios of WEM to NEM galaxies. Both Fig. 5a \&\ b show a decrease in
emission line strength as density increases, although the trend is
stronger in Fig. 5b.

Fig. 5 includes both cluster and field galaxies. Since different
processes may be operating, with different effect, inside and outside
of clusters, it is necessary to examine each population separately. We
define ``cluster galaxies'' by the method outlined in \S4. Meanwhile,
galaxies outside clusters, hereafter ``field galaxies'', are
identified by removing cluster galaxies from the entire sample, except
that this time, a lower $\Delta\rho/\rho$=40 contour is used to ensure
that galaxies in the outskirts of clusters are excluded from the field
sample. Note that the ``field'' galaxies do not necessarily consist
entirely of so called ``isolated'' galaxies, those without any
physical associations to which the galaxy belongs. Some of our field
galaxies may be members of low density associations, such as loose
groups.

Fig. 6 shows the population ratios similar with Fig. 5, but now for
the {\it field}\ sample alone. Overall, Fig. 6 still shows
qualitatively the same correlation as Fig. 5, namely, galaxies with
higher emission tend to be more abundant in less dense environments.
However, unlike Fig. 5, the SEM/WEM comparison (Fig. 6a) shows a
stronger correlation than the WEM/NEM one (Fig. 6b). In particular, in
Fig. 6b, the slope is rather flat compared to Fig. 6a (and Fig. 5b),
except for the lowest density regime. Meanwhile, Fig. 6a shows a
clear trend of stronger emission galaxies becoming more prevalent in
less dense environments.

\subsubsection{Density effects in clusters}

Fig. 7 shows the same relation as Fig. 6 for the cluster and rich
group (hereafter ``cluster'') sample, alone. Overall, again, Fig. 7
shows a qualitatively similar correlation as that in Fig. 6: galaxies
with stronger emission lines prefer less dense environments. Fig. 7a,
however, is less conclusive due to the small number of emission line
(particularly SEM) galaxies inside clusters. Meanwhile, Fig. 7b shows
a clear trend of galaxies with no emission lines becoming more
prevalent in denser environments. This is rather interesting,
especially comparing this to the previously shown (Fig. 6b) weak
correlation of WEM/NEM in the field sample. Since the WEM/NEM ratio
is expected to measure the extent of ``normal'' star formation, this
fact might suggest that the mechanism affecting normal star formation
might operate more efficiently in the cluster environment, than in the
field.

\subsection{Comparison between Clusters and Field}
 
Figures 8 shows the distribution of C for cluster and field galaxies.
The solid line represents cluster galaxies, while the dotted-dashed
line represents galaxies in the field. The C distribution of the
cluster galaxies is skewed toward early type (larger C), consistent
with the well-established trend towards larger-bulge systems inside
clusters.

Fig. 9 shows the cumulative EW(OII) distribution for the cluster and
field sample, after application of a C correction similar to that in
\S 5.3, except that weights are now assigned to field galaxies, in
order to match the C distribution of the field sample to that of the
cluster sample. The two distributions in Fig. 9 indicate that field
galaxies tend to have higher EW(OII) than cluster galaxies. A
Kolmogorov-Smirnov (KS) test shows that the probability that the two
distributions are drawn from the same parent distribution is only 5
$\times 10^{-21}$.

Fig. 9 includes galaxies of all ``structural'' types. Since the
processes leading to the cluster/field differences exhibited here
might operate differently on different type galaxies, we split the
sample into three C subclasses. Fig. 10 shows the same plot as Fig. 9
for the three separate C bins, 0.35 $< $ C $\leq $ 0.5, 0.25 $< $ C
$\leq $ 0.35, and 0.1 $< $ C $\leq $ 0.25. (The field/cluster C
correction is applied within each subclass.) Though the effect is
somewhat weaker than in Fig. 9, all three bins of Fig. 10 still show
the same trend, namely that field galaxies tend to show higher [OII]
emission than cluster galaxies. KS probabilities for the three bins
are 3 $\times 10^{-9}$, 1 $\times 10^{-14}$, and 4 $\times 10^{-3}$,
respectively. We do not, however, find any strong qualitative
differences among the three C bins.

A different look at the same trends is presented in Table 1, which
lists the percentiles of the various emission classes in different
environments. The emission classes are defined in the same way as in
\S 5.1, except that the SEM class is further subdivided into a
low subset (LSEM) and a high subset (HSEM) at the border EW(OII)= 50 \AA. 
The C corrections have {\it not} been applied in Table 1. Cluster and field
definitions are the same as in \S 4.2 and \S 5.3. Additionally, rich
and poor cluster subsets are introduced, defined by their total
luminosities $L_{T}$, which have been calculated as described in
Tucker (1994). Rich clusters are defined as clusters with $L_{T}\
\geq\ 5\times10^{11}L_{\odot} $, while poor clusters are clusters with
$L_{T} \leq\ 0.5\times10^{11}L_{\odot}$. Also included in Table 1 are 
1 $\sigma$ uncertainties, calculated from Poisson statistics.
The numbers in parentheses are the total number of galaxies in 
each environmental class.
 
Table 2 repeats the analysis of Table 1, but with C corrections. The
corrections were made using the same method as in \S 5.3, now applied
to match the C distributions of the field, rich clusters, and
all-cluster samples to the C distribution of the poor cluster
class. Although there are small differences between the numbers in the
two tables, the same trends are apparent. As was evident from Figures
9 and 10, star formation rates are higher in the field than in
clusters. However, dividing clusters into rich and poor systems
reveals some remarkable complexities underlying this general
trend. Rich clusters show a somewhat depressed level of ``normal''
star formation, as counted by the WEM fraction: a factor of 2 relative to 
field galaxies, and a factor of 1.4 when C corrections are
made. However, the frequency of starbursts, as counted by the LSEM and
HSEM fractions, is depressed by much more: a factor of about 5
relative to the field.
 
Most remarkable are the percentiles in the poor clusters. Poor
clusters show {\em higher} levels of star formation than even the
field. This enhanced star formation is particularly evident at the
highest star formation rates: the HSEM galaxies are almost 4 times
more abundant in poor clusters than in any other population. A
$\chi^{2}$\ test shows that the proportions for the poor cluster
galaxies and the field galaxies are significantly different at
significance level $\alpha$\ = 2$\times10^{-4}$ for the HSEM class.

\section{DISCUSSION}

The results of this study can be summarized as follows:

(1) The correlation between the two fundamental physical parameters underlying
the Hubble sequence, star formation rate and bulge-to-disk ratio, varies with
environment.

(2) Cluster galaxies exhibit {\em reduced} star formation compared to
the field control sample of the same concentration index, or ``galactic
structure''.
We did not find any qualitatively different responses to environments
between early and late type spirals, which some previous researches
reported.

(3) Star formation rate of galaxies of a given
``galactic structure'' is sensitive to local galaxy density and shows
a continuous correlation with the local density, in such a way that galaxies
show higher levels of star formation in low density than in high density
environments.
Remarkably, this trend is also observed both inside and outside of clusters,
implying that physical processes responsible to this
correlation might not be intrinsic to cluster environments.

(4) Among field populations, the abundance of strong emission line
galaxies, or ``starburst'' galaxies, is more sensitive to local density
than the abundance of weak emission line galaxies, i.e. ``normal star
formation'' galaxies. Among cluster populations, the opposite is true.

(5) While rich clusters show lower levels of ``normal star
formation'', and much lower levels of ``starbursts'' than the field,
poor clusters show enhanced levels of both. The starburst level in
poor clusters is a factor of four higher than that in either the field
or rich clusters.

Reviewing these results,
one might be tempted to conclude that 
star formation rates are higher in low density than in high density
environments and 
that the field/cluster difference is simply a manifestation of the
variation of SFR with local density. However, a comparison of Figures 7 and
6 shows that things are not this simple. At low levels of star
formation, the SFR is quite sensitive to density inside clusters, but
only weakly dependent on density in the field. However, this is not
the case for galaxies with high levels of star formation, which appear
to be at least as sensitive to local density in the field as in
clusters. Even this description is an oversimplification. Table 2
demonstrates that the variation of SFR with environment is not
monotonic. The highest levels of star formation are more prevelant in
the intermediate environment of poor clusters than in either the field
or rich clusters.

It is clear from these, as well as many earlier findings, that
environment has a profound effect not only on the structure but also
on the star formation rates within galaxies. Popular ideas about the
effect of environment on galactic star formation envision at least two
kinds of processes at work: those that lower the gas content, and
therefore the potential star formation rate in galaxies, and those
that precipitate bursts of star formation. Among the former are 
interactions between the intragalactic and intergalactic media,
including stripping and evaporation 
(Gunn \& Gott 1972; Cowie \& Songaila 1977), tidal
interactions which remove gas from disks (e.g. Spitzer \& Baade 1951;
Valluri \& Jog 1990), and the
suppression of infall of new gas-rich material from outside the galaxy
(Larson, Tinsley, \& Caldwell 1980). Among the latter are tidal
shocks (e.g. Noguchi \& Ishibashi 1986, Sanders et al. 1988), 
ram pressure induced star formation 
(Dressler \& Gunn 1983), and mergers with other systems 
(e.g. Barens \& Hernquist 1991).
 
Our understanding of all these processes is incomplete, and their
effects may be complex. Tidal encounters and galaxy ``harassment''
(Moore et al. 1996) might either enhance or depress average star
formation rates. More generally, short-term {\em increases} in star
formation rate will deplete the gas supply at a higher rate, leading,
perhaps, to longer-term {\em decreases} in star
formation. Nevertheless, there are some generalizations which it is
probably safe to make. Ram pressure and evaporative stripping of gas
is a process which depends on a dense and/or hot intergalactic medium,
and therefore only works well only in rich clusters. Mergers of
gas-rich systems, which probably produce starbursts, depend on the
galaxy-galaxy encounter rate. Encounters between galaxies will be more
prevalent in denser, and higher velocity dispersion environments, but
such encounters will only lead to mergers if the relative velocities
of the galaxies are comparable to or lower than the characteristic
velocities within the galaxies.
 
One can combine these generalizations with our previous inferences
that the WEM galaxies are undergoing ``normal'' star formation and the
SEM galaxies are undergoing starbursts to produce a picture which is
facile and oversimplified but may be basically true. This picture
predicts that the WEM/NEM ratio, measuring normal star formation,
should decline with density, particularly in rich clusters, which it
does. It also predicts that the SEM fraction, measuring starbursts,
should increase with density until the local velocity dispersion
exceeds internal galaxy velocities, after which it should drop. As a
result, groups and poor clusters should have the highest proportion of
SEM galaxies, which they do.
 
This may be too easy a solution. A rough correspondence of
expectations and observed trends does not prove that gas stripping and
encounter-driven starbursts are responsible for the environmental
effects on star formation rates that we observe. However, whatever the
real processes at work, we {\em can} confidently conclude that they
number at least two: one which suppresses star formation in clusters,
and one which precipitates starbursts in intermediate density
environments. Thus, one of the two fundamental galactic parameters,
star formation rate, is profoundly affected by galaxies' environments.
To what extent the other fundamental parameter, structure, is also a
product of environment will the be subject of following papers.
 
\acknowledgments

We thank Richard Larson and Richard Ellis for their useful suggestions and
Ian Smail for kindly providing his codes. 
YH acknowledge Becky Koopmann for helping the manuscript.
This work was partially supported by NSF grant AST91-15446. The Las
Campanas Redshift Survey was supported by NSF grants AST87-17207,
AST89-21326, and AST92-20460. YH was partially supported by a Carnegie
Predoctoral Fellowship.

\clearpage
\figcaption[fig1.ps]{
The distribution of equivalent widths of 15749 LCRS galaxies in the
[OII]$ \lambda$3727 versus H$\beta$ plane, where the emission is represented
by positive values.
\label{fig1}}
 
\figcaption[fig2.ps]{Optical spectra of galaxies with strong
emission lines. (a) Typical strong star-forming galaxies
(b) Typical spectrum of Seyfert 1. (c) Typical
spectrum of Seyfert2/LINER.
Wavelength is in the observed frame.
\label{fig2}}
 
\figcaption[fig3.ps]{
Relationship between the concentration index (C; Abraham et al. 1994) and mean
equivalent width of [OII]$\lambda$3727 for 16377 LCRS galaxies.
It shows a smooth increase in the mean
of EW with decreasing C, 
which parallels the relation between the
Hubble type and EW(H$\alpha$)(e.g. Kennicutt \& Kent 1983), or
EW(OII)(Kennicutt 1992)
\label{fig3}}
 
\figcaption[fig4.ps]{
Distribution of the concentration index, C in
the three emission classes; SEM (EW(OII) $\geq$ 20 \AA),
WEM (5 \AA\ $\leq $ EW(OII) $< $ 20 \AA ),
and NEM (EW(OII) $<$ 5 \AA) as defined in \S 5.3
\label{fig4}}
 
\figcaption[fig5.ps]{
Correlation between the local space density and
the population of different emission classes for LCRS galaxies.
The difference in the morphological distributions
between the emission classes are corrected using the concentration
index, C.
(a) SEM/WEM population ratio as a function of
the local space density. (b) WEM/NEM population ratio as a function of
the local space density. Bars are root N error.
\label{fig5}}
 
\figcaption[fig6.ps]{
Correlation between the local space density and
the population of different emission classes for {\it field} subsets.
The C correction is also applied.
\label{fig6}}
 
\figcaption[fig7.ps]{
Correlation between the local space density and
the population of different emission classes for {\it cluster} subsets,
after the C correction.
\label{fig7}}

\figcaption[fig8.ps]{
Distribution of the concentration index C
of the cluster galaxies and the field galaxies.
The solid line represents cluster galaxies,
while the dotted-dashed line represents galaxies in the field.
The C distribution of the
cluster galaxies is skewed toward early type (larger C), consistent
with the well-established trend towards larger-bulge systems inside
clusters.
\label{fig8}}
 
\figcaption[fig9.ps]{
Fig. 9 shows the cumulative EW(OII) distribution for the cluster and
field sample, after application of the C correction.
The two distributions indicate that field
galaxies tend to have higher EW(OII) than cluster galaxies.
\label{fig9}}
 
\figcaption[fig10.ps]{
The same plot as Fig. 9 for three separate C bins,
0.35 $< $ C $\leq $ 0.5, 0.25 $< $ C $\leq $ 0.35, and 0.1 $< $ C $\leq $
0.25. (The C correction is also applied.) 
Though the effect is somewhat weaker than in Fig. 9,
all three bins of Fig. 10 still show the same trend, namely that field 
galaxies tend to show higher [OII] emission than cluster galaxies.
\label{fig10}}

\clearpage
\begin{deluxetable}{crrrrrrrrrrr}
\footnotesize
\tablenum{1}
\tablecaption{Percentiles of Emission Classes}
\tablewidth{0pt}
\tablehead{
\colhead{Class} & \colhead{Field} & \colhead{Cluster} & \colhead{Cluster } & \colhead{Cluster}\\
\colhead{ } & \colhead{ } & \colhead{Poor} & \colhead{All} & \colhead{Rich} \\
\colhead{ } & \colhead{(6051)} & \colhead{(346)} & \colhead{(3825)} & \colhead{(394)}
}
\startdata
NEM & 56.8$\pm$0.6 & 43.6$\pm$2.7 & 68.7$\pm$0.8 & 80.5$\pm$2.0 \nl
WEM & 33.1$\pm$0.6 & 39.6$\pm$2.6 & 24.6$\pm$0.7 & 17.6$\pm$1.9 \nl
LSEM & 9.1$\pm$0.3 & 13.8$\pm$1.8 & 5.9$\pm$0.3 & 1.5$\pm$0.6 \nl
HSEM & 0.9$\pm$0.1 & 3.0$\pm$0.8 & 0.8$\pm$0.1 & 0.4$\pm$0.3\nl

\enddata
\end{deluxetable}
\clearpage
\begin{deluxetable}{crrrrrrrrrrr}
\footnotesize
\tablenum{2}
\tablecaption{Percentiles of Emission Classes (with C Correction)}
\tablewidth{0pt}
\tablehead{
\colhead{Class} & \colhead{Field} & \colhead{Cluster} & \colhead{Cluster } & \colhead{Cluster}\\
\colhead{ } & \colhead{ } & \colhead{Poor} & \colhead{All} & \colhead{Rich} \\
}
\startdata
NEM & 55.3$\pm$2.6 & 43.6$\pm$2.7 & 62.8$\pm$2.6 & 73.9$\pm$2.4 \nl
WEM & 33.5$\pm$2.5   & 39.6$\pm$2.6 & 29.2$\pm$2.4 & 23.9$\pm$2.3 \nl
LSEM & 10.3$\pm$1.6  & 13.8$\pm$1.8 & 6.9$\pm$1.4 & 1.5$\pm$0.7 \nl
HSEM & 0.8$\pm$0.4   & 3.0$\pm$0.8  & 1.0$\pm$0.5 & 0.7$\pm$0.4\nl

\enddata
\end{deluxetable}

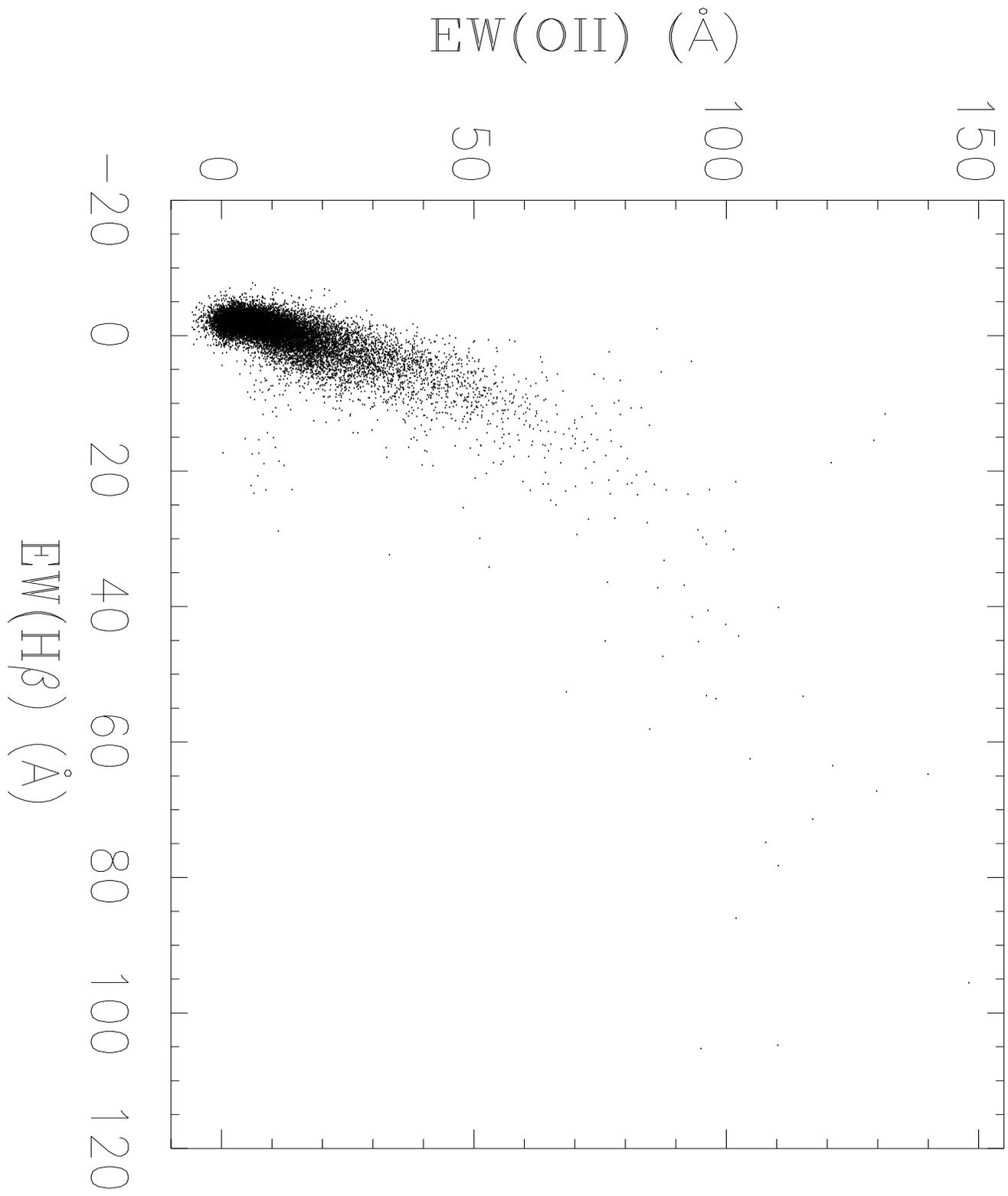
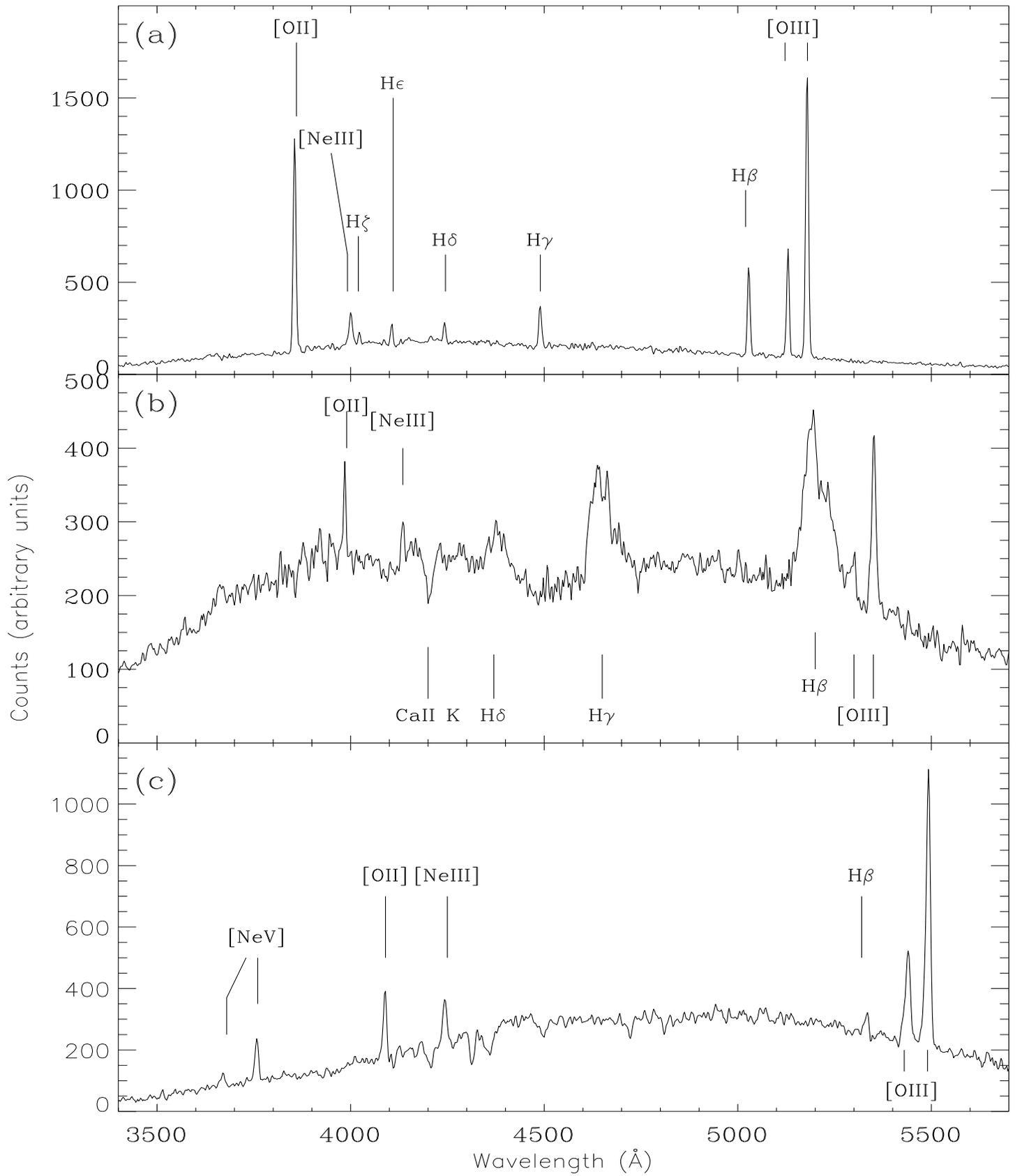
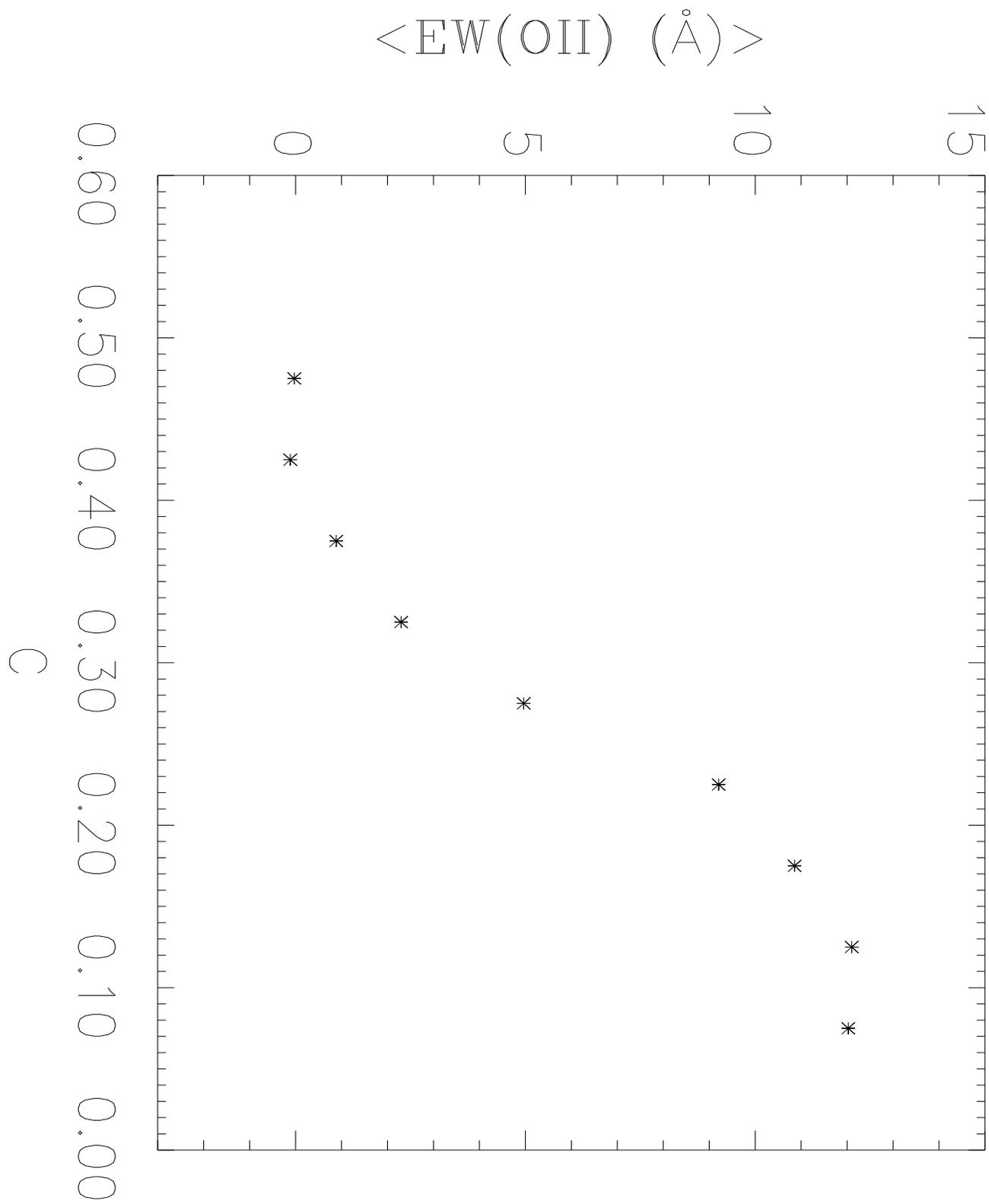
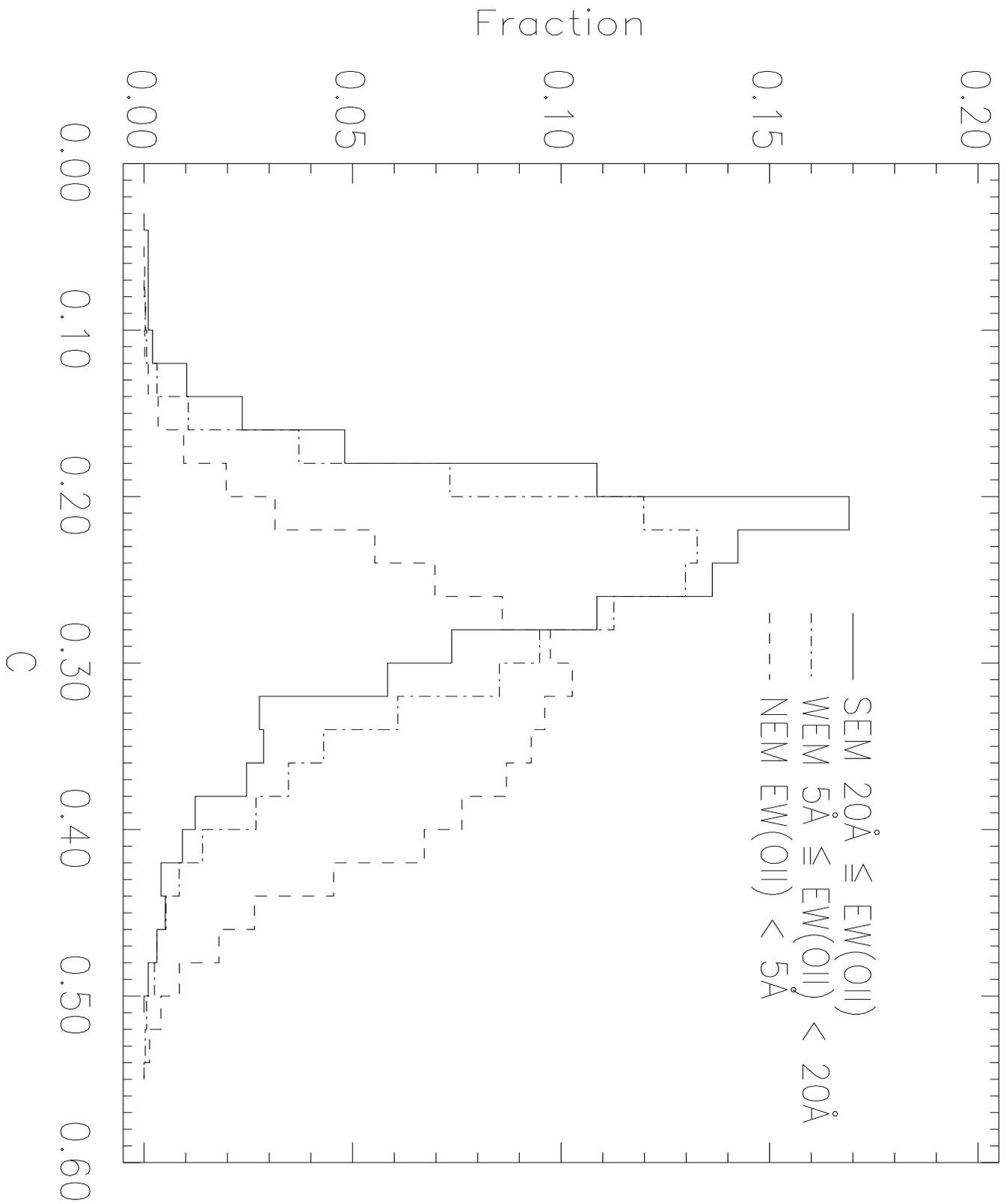
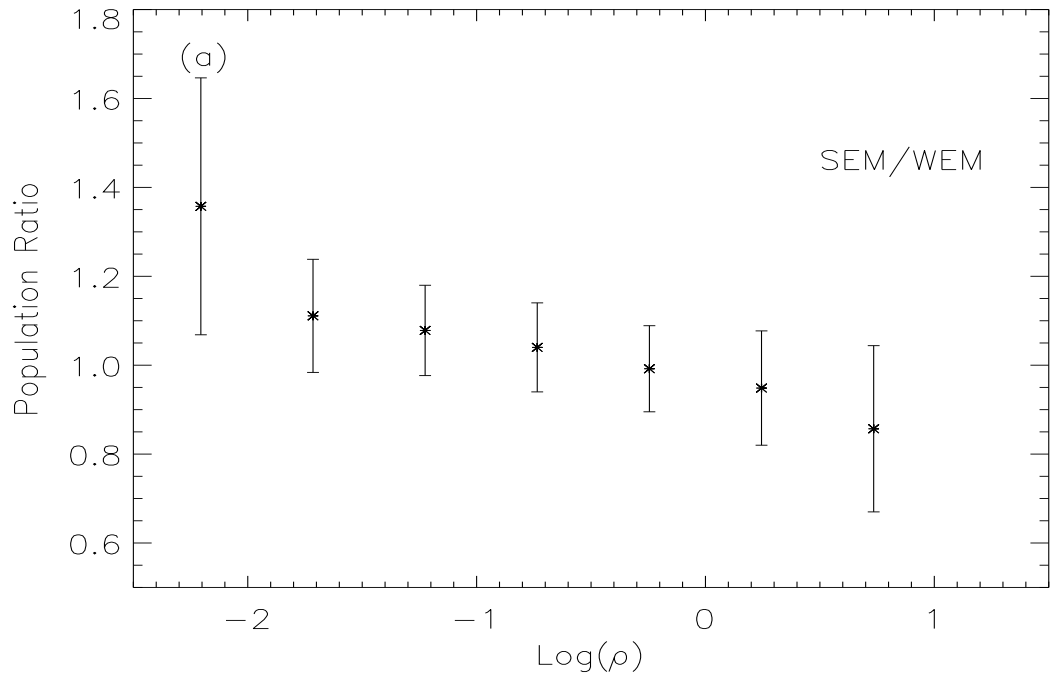
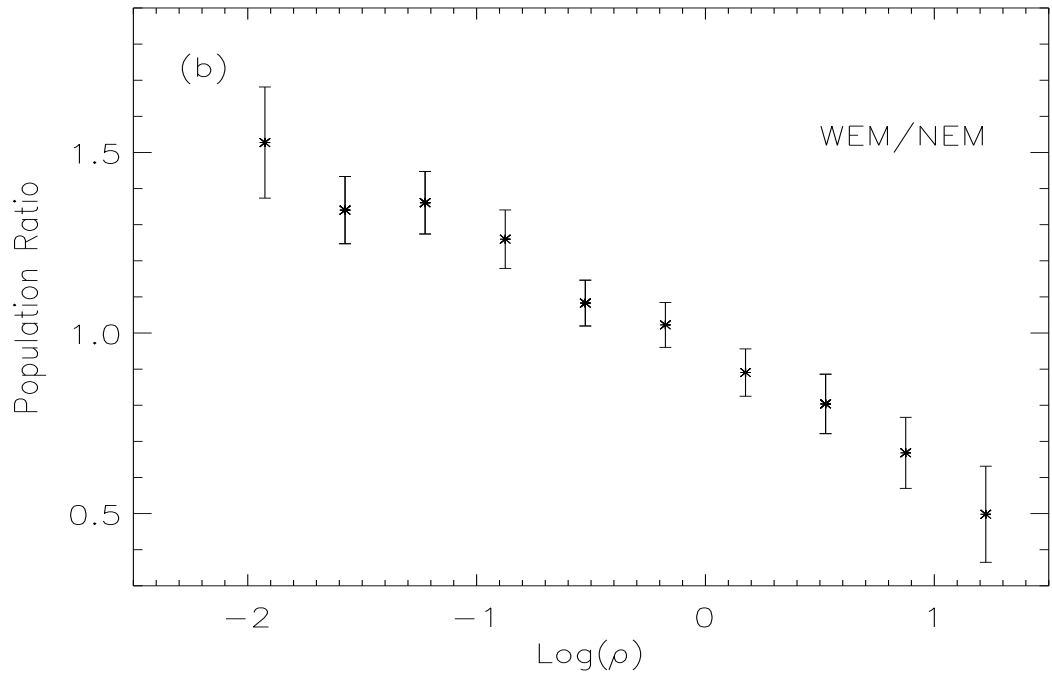
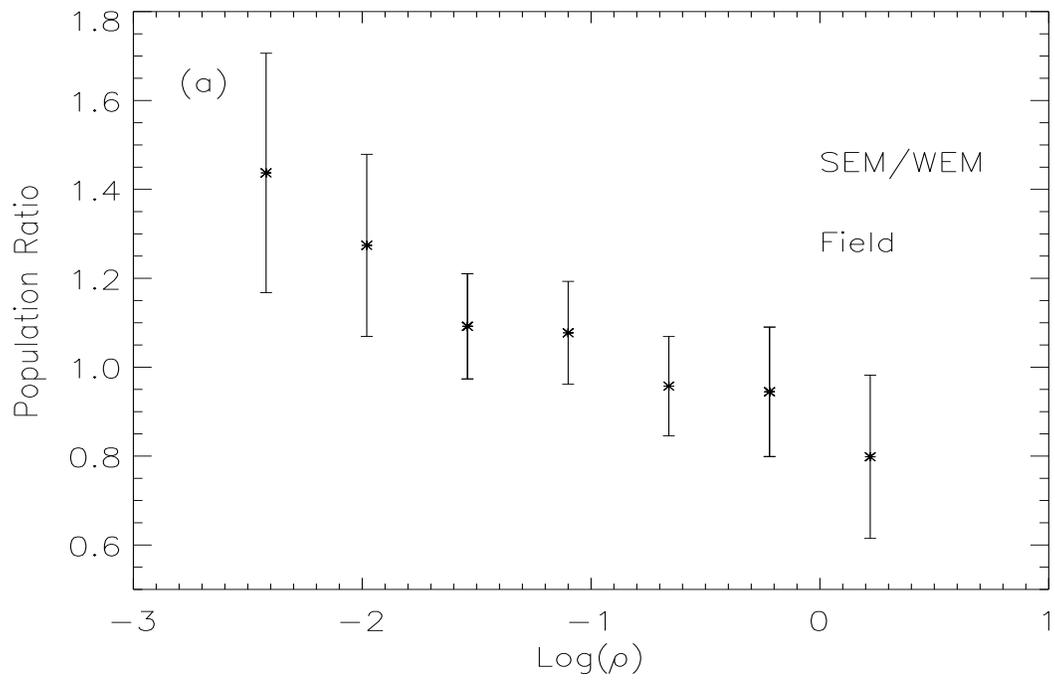
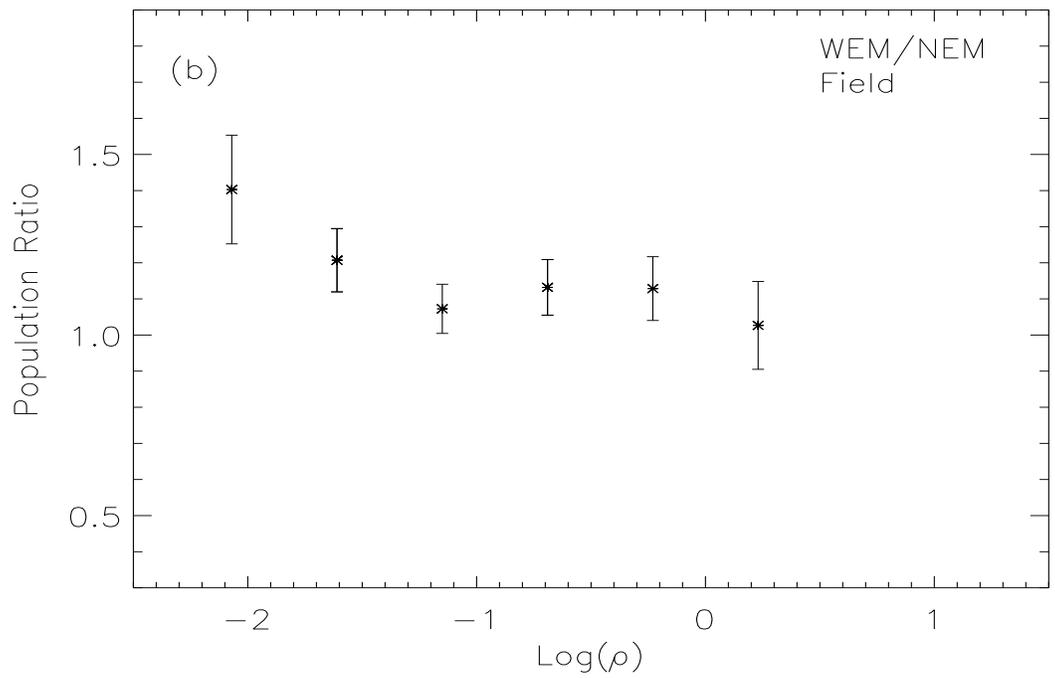
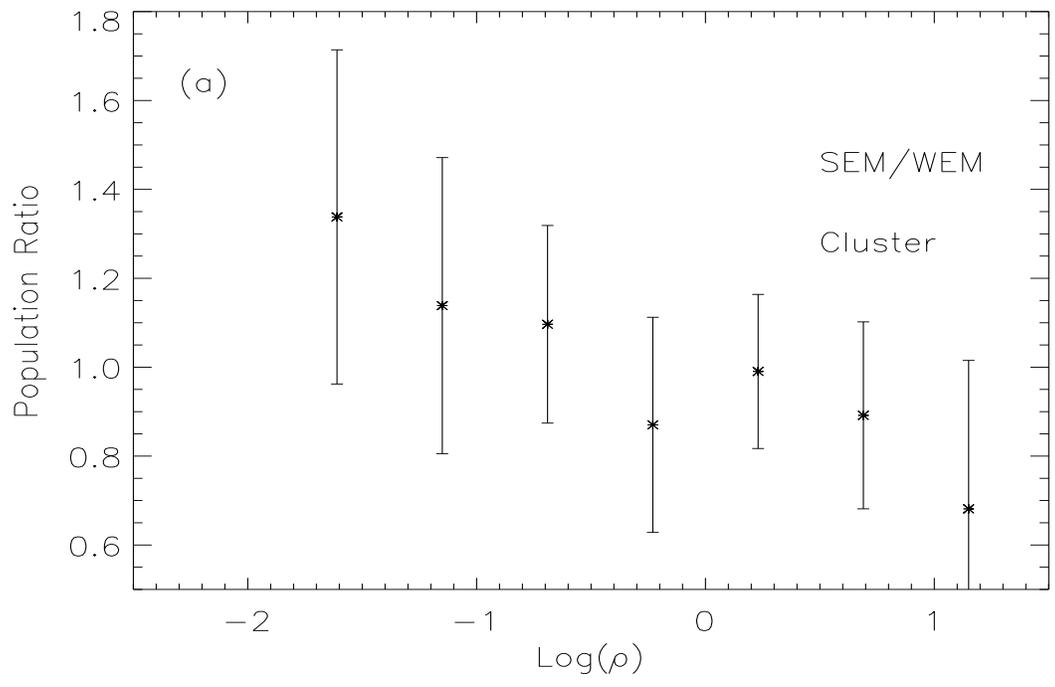
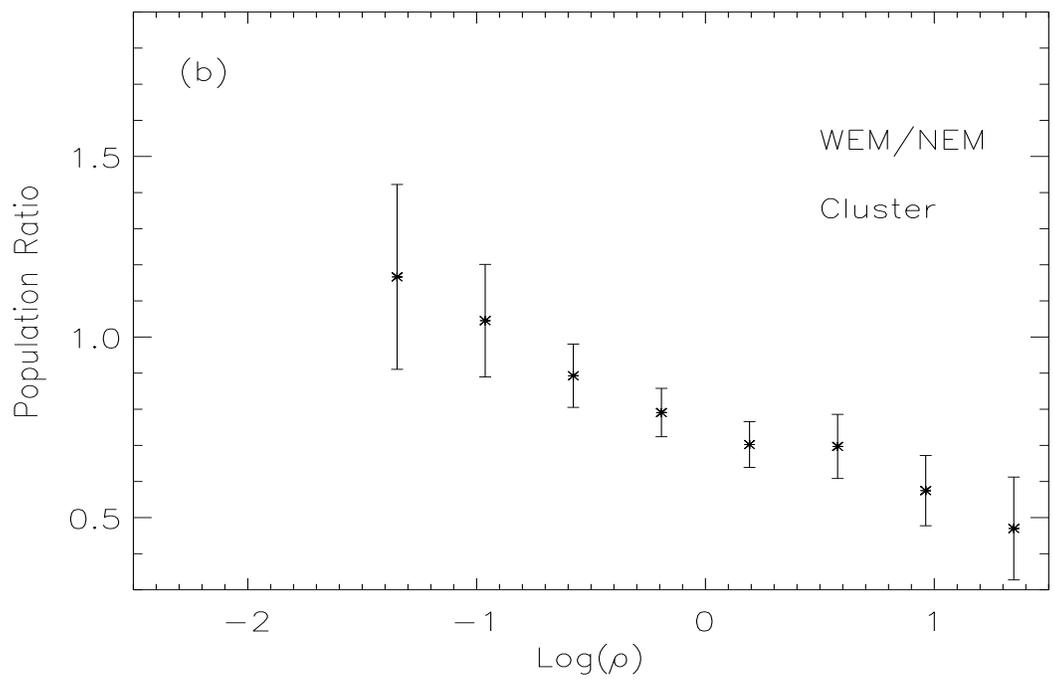
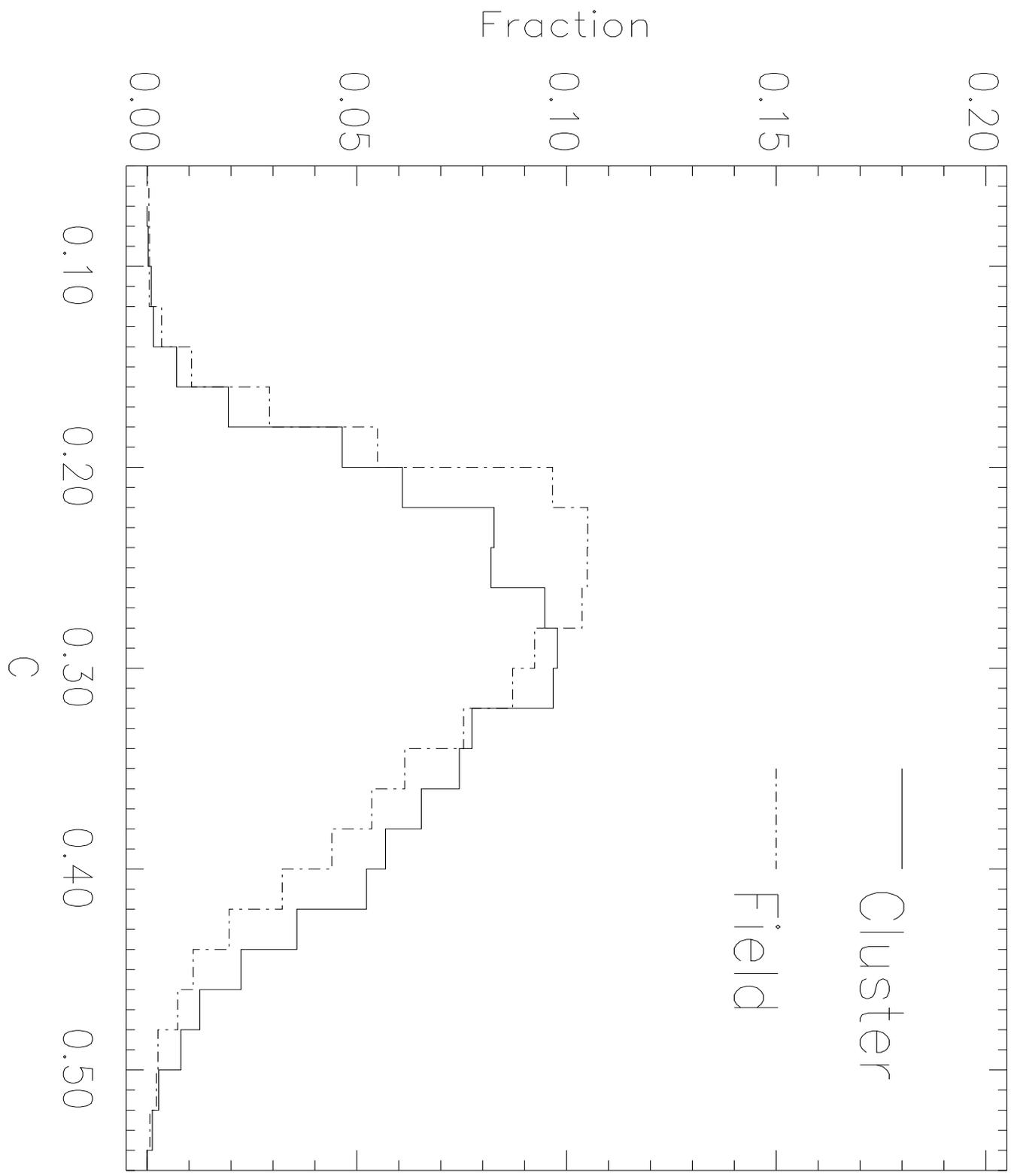
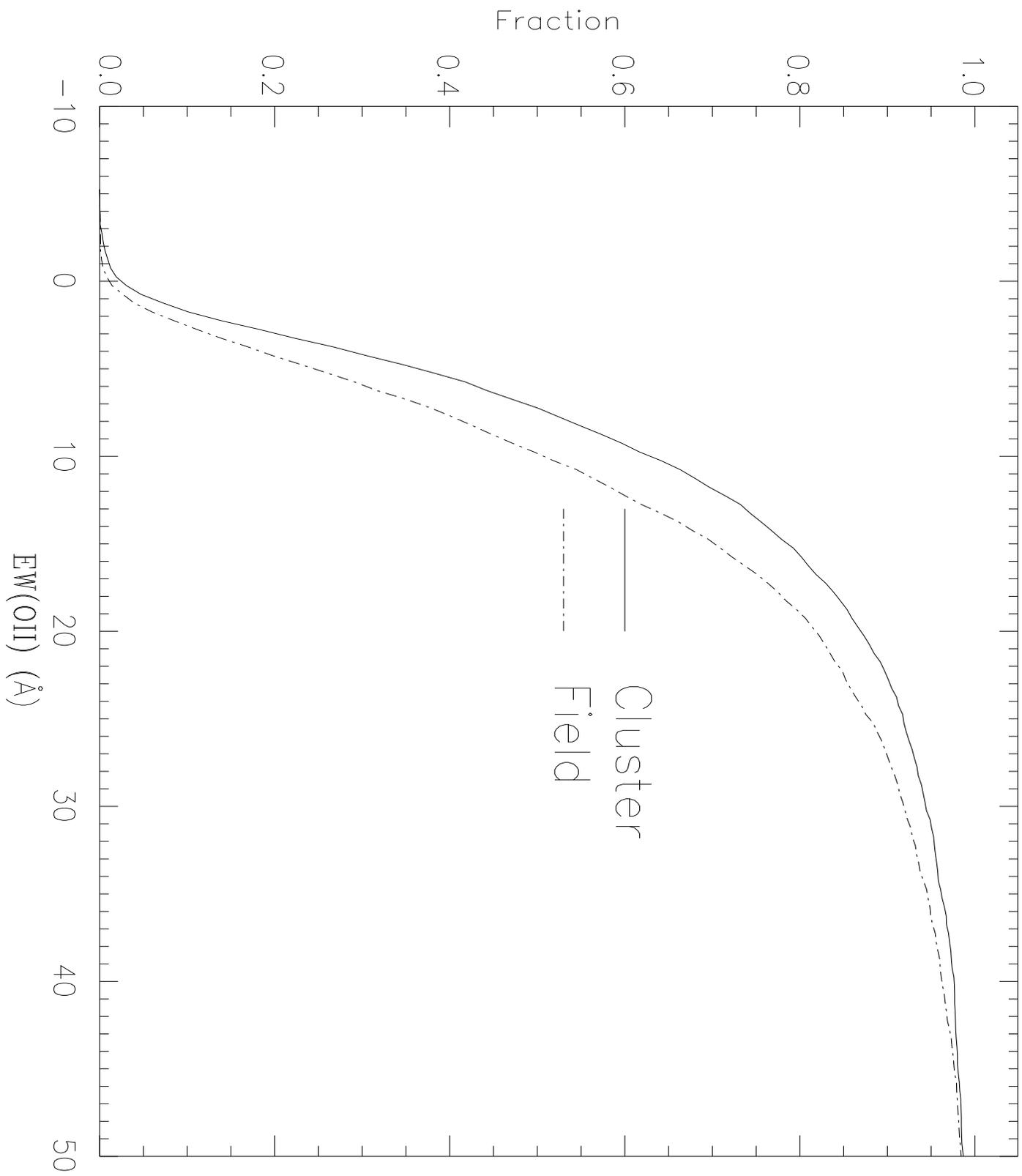
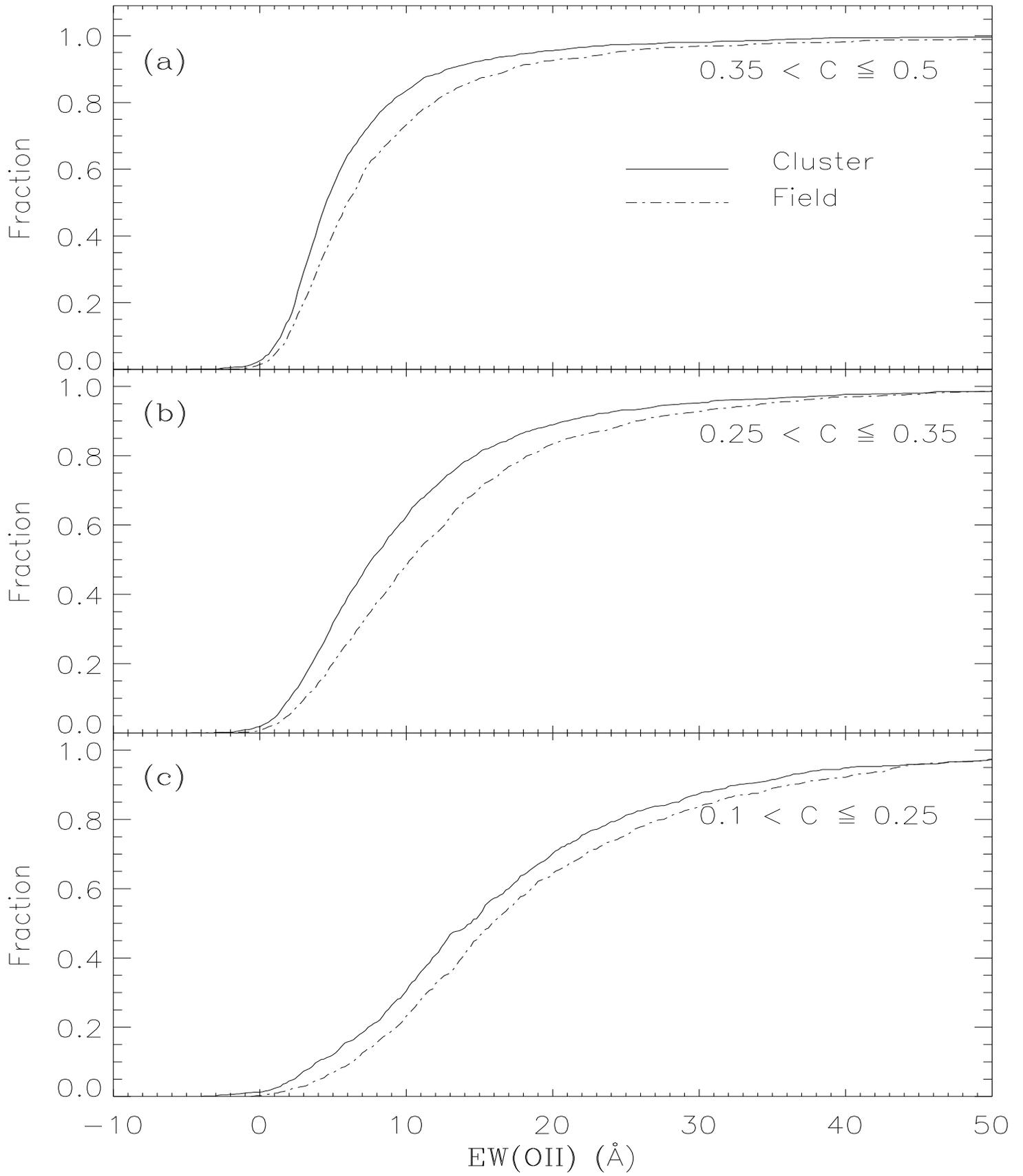
\end{document}